\documentstyle[preprint,aps,prl,epsfig]{revtex}

\tighten
\begin{document}
\draft
\preprint{version 2.0}

\title{Logarithmic two-loop corrections to the Lamb shift in hydrogen }

\author{Krzysztof Pachucki
\thanks{E-mail address: krp@fuw.edu.pl}}
\address{
Institute of Theoretical Physics, Warsaw University,
Ho\.{z}a 69, 00-681 Warsaw, Poland}
\maketitle

\begin{abstract}
Higher order $(\alpha/\pi)^2\,(Z\,\alpha)^6$ logarithmic corrections 
to the hydrogen Lamb shift are calculated. The results obtained show 
the two-loop contribution has a very peculiar behavior,
and significantly 
alter the theoretical predictions for low lying S-states.
\end{abstract}
\narrowtext

The calculation of the two-loop contribution to the Lamb shift in hydrogen 
is one of the most challenging projects in bound state QED \cite{sap1,eides2}.
Since direct numerical calculations with the use of Dirac-Coulomb
propagators have not yet been completed, one has to rely on the $Z\,\alpha$
expansion:
\begin{eqnarray}
\Delta E &=& m\,\biggl(\frac{\alpha}{\pi}\biggr)^2\,(Z\,\alpha)^4\,
\biggl\{ B_{40}+(Z\,\alpha)\,B_{50}\nonumber \\ &&
+(Z\,\alpha)^2\,\biggl[ \ln^3(Z\,\alpha)^{-2}\,B_{63}+
 \ln^2(Z\,\alpha)^{-2}\,B_{62}+
 \ln(Z\,\alpha)^{-2}\,B_{61}+ B_{60} \biggr]+\ldots \biggr\}.
\label{1}
\end{eqnarray}
The leading order correction $B_{40}$ can
be obtained from the slope of the electron  form-factors $F_1$ and $F_2$
at $q^2=0$. It is known analytically and its numerical value is 
quite small (for $S$-states including vacuum polarization)
\begin{equation}
B_{40} = 0.538941\,. \label{2}
\end{equation}
The calculation of the next order correction $B_{50}$ was completed
only a few years ago independently by two groups in \cite{krp1} and \cite{eides1}. 
The value was surprisingly large
\begin{equation}
B_{50} = -21.5561(31)\,.
\end{equation}
Moreover, this correction led to strong disagreement
in He$^+$ Lamb shift with the most precise experimental value in \cite{hel},
while for hydrogen Lamb shift, led to agreement with the  Mainz value for the 
proton charge radius \cite{krp2}. 
This large value of $B_{50}$ compared to $B_{40}$
indicates a~very slow convergence or even 
might suggest a nonperturbative behavior of the two-loop contribution.
Indeed, the direct numerical calculations of one diagram,
the loop-by-loop electron self--energy by Mallampalli and Sapirstein in \cite{sap2}, 
shows that the value of this correction at $Z=1$ is of different sign
and magnitude, that the one based on the first two terms of
analytic expansion. Moreover, this numerical calculation
was in disagreement with the analytical value of $B_{63}$ in \cite{sk1}, while 
it was argued in \cite{sk1}, that this correction comes only from this one diagram
in the covariant gauge. A year later another group \cite{soff} calculated
numerically this one diagram and found agreement with 
the analytic expansion including $\ln^3(Z\,\alpha)^{-2}$ term.
While this may suggest that the first numerical calculations were incorrect,
a very recent, third  numerical result by Yerokhin in \cite{yer} 
confirmed the first one \cite{sap2}.
So, this situation with the two-loop contribution is very unclear.
Moreover, the analytic value  $\ln^3(Z\,\alpha)^{-2}$ term 
corresponding to all diagrams, was 
confirmed independently by several groups, so this situation is even more
confusing. It was argued, by Yerokhin in \cite{yer}, that the  
$\ln^3(Z\,\alpha)^{-2}$ term for this one loop-by-loop diagram 
is different from the total value of $B_{63}$, and in fact 
found an additional contribution. 
However, the value for this term, coming from all diagrams
might be correct, because other diagrams may contain compensating terms.   
The goal of this work is to shed some light into higher order
two-loop corrections and calculate all logarithmic terms:
$B_{63}, B_{62}$, and $B_{61}$. We find that indeed the two-loop contribution 
has a very peculiar behavior, as the higher order term $B_{61}$ dominates
and reverses the sign for the overall logarithmic
contribution. In next sections
we present some details of this calculation. First, a simple example
is worked out to demonstrate the method, then we pass to the most difficult
two-photon-loop diagrams and complete with remaining diagrams
containing an electron loop. Conclusions with prospects of calculation
of $B_{60}$ summarize this work.   

\section{Simple example}
The example to demonstrate the calculational method is 
the asymptotic expansion of
\begin{eqnarray}
P(\omega) &\equiv& \langle\phi|\,\bbox{p}\,\frac{1}{E-(H+\omega)}\,
\bbox{p}\,|\phi\rangle \\
 &=& - \frac{1}{\omega}+ \frac{2}{\omega^2}-
   \frac{4\,{\sqrt{2}}}{\omega^{\frac{5}{2}}}+ 
   \frac{4 - 12\,\ln (2) + 4\,\ln (\omega)}{\omega^3}+\ldots
\label{5}
\end{eqnarray}
around large $\omega$ for ground state of the hydrogen atom. 
More precisely, we concentrate on the $\omega^{-3}$ term. 
For simplicity, we put here $m=1, \alpha=1$.
From one side $P(\omega)$ is known analytically \cite{gaw}
\begin{equation}
P(\omega) = -\frac{384\, \tau^5}{(1+\tau)^8\, (2-\tau)} \,
            {}_2F_1(4,2-\tau,3-\tau,\zeta)\,, \label{6}
\end{equation}
where
\begin{eqnarray}
\zeta = \biggl(\frac{1-\tau}{1+\tau}\biggr)^2\,,\;\;
\tau = \frac{1}{\sqrt{2\,(\omega+1/2)}}\,,
\end{eqnarray}
so one could get this coefficient from here. However, our final 
goal is to calculate the two-loop contribution, for which no analytic formula
has been derived so far. Therefore, we use a different approach, based
on the effective Hamiltonian. First, we regularize the Coulomb interaction
by the following replacement:
\begin{equation}
V(r) = -\frac{1}{r} \rightarrow -\frac{1}{r}\,(1-e^{-\lambda\,r})\,.
\label{8}
\end{equation}
With the regularized potential ($P\rightarrow P_R$)
one can expand $P_R$ in $(H-E)/\omega$ which leads to
\begin{eqnarray}
P_R &=& -\frac{1}{\omega^3}\,
\langle\phi|\bbox{p}\,(H-E)^2\,\bbox{p}|\phi\rangle
= -\frac{1}{\omega^3}\,\langle\phi|V'(r)^2|\phi\rangle\,, \\
\langle\phi|V'(r)^2|\phi\rangle &=& 2\,\lambda+8\,\ln(3)-8\,\ln(\lambda)-2\,,
\end{eqnarray} 
where the last expectation value is taken from \cite{krp3}. The remaining part,
which was left out by this replacement, is obtained from the subtracted
forward scattering amplitude. Two photon exchange is
\begin{equation}
P_2 = \int\frac{d^3p}{(2\,\pi)^3}\,64\,\pi\,\biggl[\,
\frac{\bbox{p}}{p^4}\,\frac{(-1)}{p^2/2+\omega}\,\frac{\bbox{p}}{p^4}
-\frac{\bbox{p}}{p^4}\,\frac{\lambda^2}{p^2+\lambda^2}\,
\frac{(-1)}{p^2/2+\omega}\,\frac{\lambda^2}{p^2+\lambda^2}
\,\frac{\bbox{p}}{p^4}\biggr] = \frac{2\,\lambda}{\omega^3}\,,
\end{equation}
where we keep only the $\omega^{-3}$ term ($\omega^{-1}$ and $\omega^{-2}$
are subtracted out before the integration).
The three photon exchange requires more subtractions.
One Coulomb exchange between photon vertices gives $P_{3A}$ 
\begin{eqnarray}
P_{3A} &=&\int\frac{d^3p}{(2\,\pi)^3}\,
\int\frac{d^3p'}{(2\,\pi)^3}
\,64\,\pi\,\biggl[\,
\frac{\bbox{p}'}{p'^4}\,\frac{(-1)}{p'^2/2+\omega}\,
\frac{(-4\,\pi)}{q^2}\,\frac{(-1)}{p^2/2+\omega}\,\frac{\bbox{p}}{p^4}
\nonumber \\ &&
-\frac{\bbox{p}'}{p'^4}\,\frac{\lambda^2}{p'^2+\lambda^2}\,
\frac{(-1)}{p'^2/2+\omega}\,
\frac{(-4\,\pi)}{q^2}\,\frac{\lambda^2}{\lambda^2+q^2}\,
\frac{(-1)}{p^2/2+\omega}\,
\frac{\lambda^2}{p^2+\lambda^2}
\,\frac{\bbox{p}}{p^4}\biggr] \nonumber \\
&=& \frac{4\,\ln\omega-8\,\ln\lambda-8\,\ln3+20\,\ln2}{\omega^3}\,.
\end{eqnarray}
Coulomb exchanges out of photon vertices gives $P_{3B}$
\begin{eqnarray}
P_{3B} &=& -2048\,\pi^2\,
\int\frac{d^3p}{(2\,\pi)^3}\,
\int\frac{d^3p'}{(2\,\pi)^3}\,\biggl(
\frac{1}{p'^4}\,\frac{1}{q^2}\,\frac{1}{p^2+2\,\omega}\,\frac{1}{p^4}
\nonumber \\ &&
-\frac{1}{p'^4}\,\frac{\lambda^2}{\lambda^2+p'^2}\,
\frac{1}{q^2}\,\frac{\lambda^2}{\lambda^2+q^2}\,
\frac{1}{p^2+2\,\omega}\,\frac{\lambda^2}{\lambda^2+p^2}\,\frac{1}{p^4}
\biggr)\nonumber \\
 &=& \frac{2-32\,\ln(2)+16\,\ln(3)}{\omega^3}\,.
\end{eqnarray}
There is an implicit subtraction at $p'=0$ for removal of small $p'$
divergence. It corresponds to subtraction of lower order contributions.
Additionally, only the $\omega^{-3}$ term is selected. The sum
\begin{equation} 
P = P_R+P_2+P_{3A}+P_{3B} = \frac{4 - 12\,\ln (2) + 4\,\ln (\omega)}{\omega^3}
\end{equation}
is independent of $\lambda$ in the limit of large $\lambda$
and agrees with that from the expansion of analytic formula in Eq. (\ref{5}).
The advantage of this method is the direct application to
the two-loop Lamb shift.

\section{Two-loop Lamb shift}
The calculations of the two-loop Lamb shift in the order of $\alpha^2(Z\,\alpha)^6$
is more complicated due to the presence of powers of $\ln(Z\,\alpha)$.
It reflects the fact that several energy and momentum regions contribute.
For these calculations we introduce a number of cutoff parameters to separate
different regions and calculate them independently. In Fig. \ref{fig1}
the integration region of two photon energies $\omega_1$ and $\omega_2$
is split with the help of $\epsilon_1,\epsilon_2, \epsilon'_1,\epsilon'_2$.
Additionally $\lambda$ 'splits' the integration over electron momenta.
The splitting itself, does not help too much. The key trick is
the assumption that after expansion in $Z\,\alpha$ one goes to the limits
$\epsilon_2\rightarrow 0,
 \epsilon_1\rightarrow 0, 
 \epsilon'_2\rightarrow 0,
 \epsilon'_1\rightarrow 0,
 \lambda\rightarrow \infty$, in the order as written.
The two-loop contribution is split accordingly
\begin{equation}
\Delta E = E_L+E_M+E_F+E_H\,,
\end{equation}
and calculated separately, each term in the most convenient gauge.
In the following sections we calculate all logs. The constant term $B_{60}$
is left unevaluated, however we lay the groundwork for its calculation.

\section{Contribution $E_L$}
\begin{figure}
\centerline{\psfig{figure=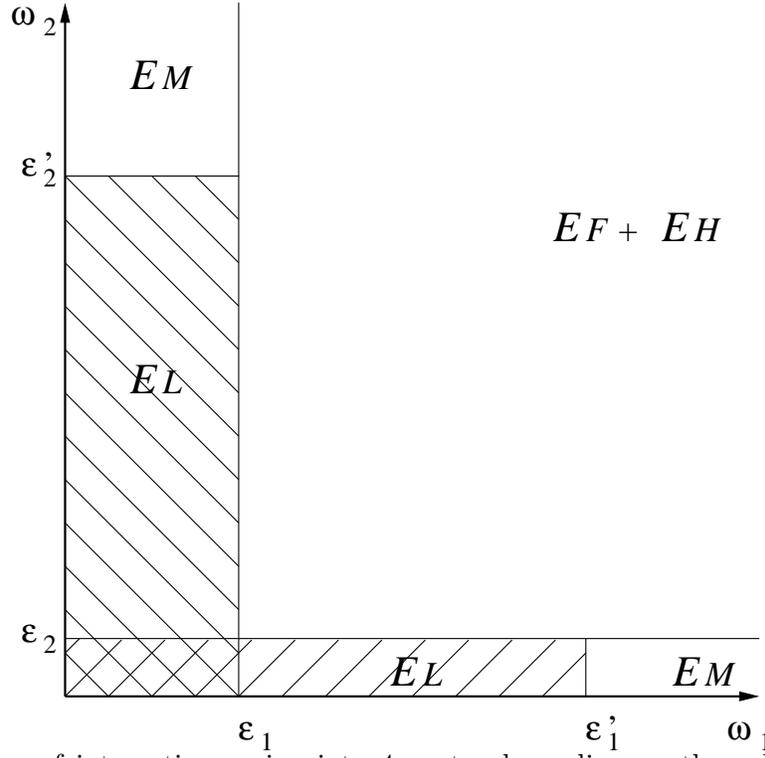,width=4in}}
 \caption{division of integration region into 4 parts, depending on the value
          of both photon frequencies, $\epsilon_2 << \epsilon_1$}
\label{fig1}
\end{figure}

\begin{figure}
\centerline{\psfig{figure=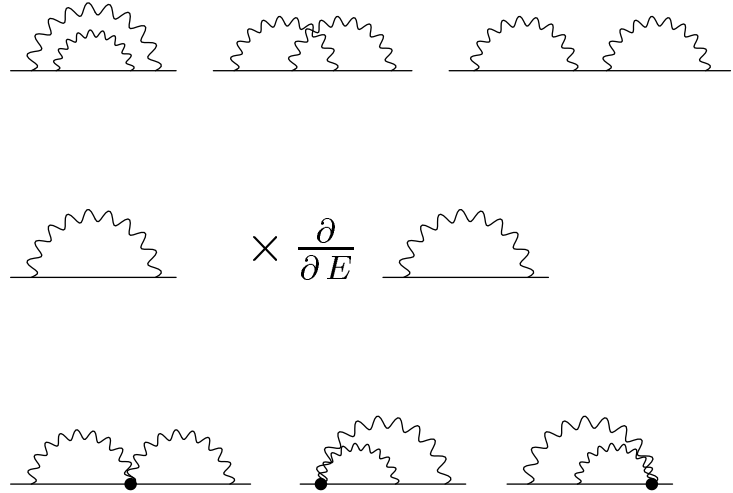,width=4in}}
 \caption{Two--loop diagrams in the Coulomb gauge in NRQED}
\label{fig2}
\end{figure}

Diagrams in the Coulomb gauge in NRQED are presented in Fig. \ref{fig2}.
We calculate them first, for photon energies inside a rectangular box
$0<\omega_1<\epsilon_1, 0<\omega_2<\epsilon_2, \epsilon_2<<\epsilon_1$,
and after combine to the region $E_L$ as shown in Fig. \ref{fig1}.
The expression derived from nonrelativistic QED for all these diagrams is:
\begin{eqnarray}
{\cal E}_L &=&\left(\frac{2\,\alpha}{3\,\pi\,m^2}\right)^2\,
\int_0^{\epsilon_1}\,d\omega_1\,\omega_1\,
\int_0^{\epsilon_2}\,d\omega_2\,\omega_2
 \nonumber \\ && \biggl\{
\langle\phi|p^i\,\frac{1}{E-(H+\omega_1)}\,p^j\,
\frac{1}{E-(H+\omega_1+\omega_2)}\,p^i\,\frac{1}{E-(H+\omega_2)}\,
p^j|\phi\rangle \nonumber \\
&&+\frac{1}{2}\,\langle\phi|p^i\,\frac{1}{E-(H+\omega_1)}\,p^j\,
\frac{1}{E-(H+\omega_1+\omega_2)}\,p^j\,\frac{1}{E-(H+\omega_1)}\,
p^i|\phi\rangle \nonumber \\ &&
+\frac{1}{2}\,\langle\phi|p^i\,\frac{1}{E-(H+\omega_2)}\,p^j\,
\frac{1}{E-(H+\omega_1+\omega_2)}\,p^j\,\frac{1}{E-(H+\omega_2)}\,
p^i|\phi\rangle \nonumber \\&& +
\langle\phi|p^i\,\frac{1}{E-(H+\omega_1)}\,p^i\,
\frac{1}{(E-H)'}\,p^j\,\frac{1}{E-(H+\omega_2)}\,
p^j|\phi\rangle \nonumber \\&& -
\frac{1}{2}\,\langle\phi|p^i\,\frac{1}{E-(H+\omega_1)}\,p^i|\phi\rangle\,
\langle\phi|p^j\,\frac{1}{[E-(H+\omega_2)]^2}\,p^j|\phi\rangle 
\nonumber \\ &&-
\frac{1}{2}\,\langle\phi|p^i\,\frac{1}{E-(H+\omega_2)}\,p^i|\phi\rangle\,
\langle\phi|p^j\,\frac{1}{[E-(H+\omega_1)]^2}\,p^j|\phi\rangle
\nonumber \\ &&+
m\,\langle\phi|p^i\,\frac{1}{E-(H+\omega_1)}\,
\frac{1}{E-(H+\omega_2)}\,p^i|\phi\rangle\nonumber \\ &&-
\frac{m}{\omega_1+\omega_2}\langle\phi|p^i\,\frac{1}{E-(H+\omega_2)}\,
p^i|\phi\rangle-
\frac{m}{\omega_1+\omega_2}\langle\phi|p^i\,\frac{1}{E-(H+\omega_1)}\,
p^i|\phi\rangle\biggr\}\,. \label{16}
\end{eqnarray}
It is a two-loop analog of Bethe logs. We have not found a way to calculate
its matrix elements analytically in a compact form, therefore we proceed in
a different way. One finds, that ${\cal E}_L$ as in Eq. (\ref{16}) depends
on $\alpha$ only through $\epsilon_1$ and $\epsilon_2$: 
\begin{equation}
{\cal E}_L = {\cal E}_L\Bigl(\frac{\epsilon_1}{\alpha^2},
                             \frac{\epsilon_2}{\alpha^2}\Bigr)\,.
\end{equation}
To find the logarithmic dependence, we differentiate ${\cal E}_L$
over $\epsilon_1$ and $\epsilon_2$ which with the help of 
$\epsilon_2<<\epsilon_1$  leads to a much simpler expression.
The first derivative leads to
\begin{eqnarray}
\epsilon_1\,\frac{\partial {\cal E}_L}{\partial \epsilon_1} &=&
\biggl(\frac{2\,\alpha}{3\,\pi\,m^2}\biggr)^2\,
\int_0^{\epsilon_2}d\omega_2\,\omega_2\,
\delta_{\pi\,\delta^3(r)}\,
\langle\phi|p^i\,\frac{1}{E-(H+\omega_2)}\,p^i|\phi\rangle\,,
\end{eqnarray} 
where $\delta_{\pi\,\delta^3(r)}$ denotes first order corrections
to $\phi, H,E$ due to $\pi\,\delta^3(r)$ operator. This integral was considered
and calculated in the context of hyperfine splitting in hydrogen-like 
systems \cite{krp4}, since the Fermi spin-spin interaction is also 
proportional to $\delta^3(r)$.
The result from that paper which is extended here to any value
of principal quantum number is:
\begin{equation}
\frac{2\,\alpha}{3\,\pi\,m^2}\,\delta_{\pi\,\delta^3(r)}\, 
\int_0^{\epsilon}d\omega\,\omega\,
\langle\phi|p^i\,\frac{1}{E-(H+\omega)}\,p^i|\phi\rangle
= \frac{\alpha}{\pi}\,\alpha^2\,\frac{F(n)}{n^3}\,,
\end{equation}
\begin{equation}
F(n) = -\frac{2}{3}\,\ln^2\bar{\epsilon} 
+\ln\bar{\epsilon}\,\biggl[2\,(1-2\,\ln(2))+\frac{8}{3}\,
\biggl(\frac{3}{4}+\frac{1}{4\,n^2}-\frac{1}{n}-\ln(n)+\Psi(n)+C
\biggr)\biggr]+N(n)\,, \label{20}
\end{equation}
where $N$ has been calculated  only for $n=1$
\begin{equation}
N \equiv N(1) = 17.8299093\,,
\end{equation}
and $\Psi = \Gamma'/\Gamma$ with Euler $\Gamma$ function and Euler $C$ constant
\begin{equation}
\Psi(1) = -C;\;\;
\Psi(n) = 1+\frac{1}{2}+\frac{1}{3}+ \ldots+\frac{1}{n-1}-C
\end{equation} 
We have introduced here a notation $\bar{\epsilon} = \epsilon/\alpha^2$,
which is to be used throughout this work. 
The result for  $n=1$ with ${\cal E} = m\,(\alpha/\pi)^2\,\alpha^6$ is:
\begin{equation}
\epsilon_1\,\frac{\partial {\cal E}_L}{\partial \epsilon_1} =
{\cal E}\,\frac{2}{3}\,
\biggl[-\frac{2}{3}\,\ln^2(\bar{\epsilon_2})
+2\,(1-2\,\ln 2)\,\ln(\bar{\epsilon_2})+N\biggr]\,.
\label{22}
\end{equation}
The second derivative, over $\epsilon_2$, is little more difficult to calculate:
\begin{eqnarray}
\epsilon_2\,\frac{\partial {\cal E}_L}{\partial \epsilon_2} &=&
\biggl(\frac{2\,\alpha}{3\,\pi\,m^2}\biggr)^2\,
\int_0^{\epsilon_1}d\omega_1\,\omega_1\,\epsilon_2^2\,\Bigl\{\ldots\Bigr\}
\nonumber \\
&=& \biggl(\frac{2\,\alpha}{3\,\pi\,m^2}\biggr)^2\,
\biggl(\int_{0}^{\epsilon'_1}+\int_{\epsilon'_1}^{\epsilon_1}\biggr)
d\omega_1\,\omega_1\,\epsilon_2^2\,\Bigl\{\ldots\Bigr\}=A+B\,. \label{24}
\end{eqnarray}
One splits it into two parts, with the assumption $\epsilon'_1 <<\epsilon_2$.
The first term $A$ has the same form as that in Eq. (\ref{22})
with $\epsilon_2$ replaced by $\epsilon'_1$. 
The second term $B$ is in turn split into two parts $B=B_L+B_H$, where
$B_L$ is calculated with the regularized Coulomb potential, as in Eq. (\ref{8}).
One can expand here in the ratio $(H-E)/\omega$  which leads to the
expression:
\begin{equation}
B_L = \frac{{\cal E}}{9}\,
\ln\biggl(\frac{\bar{\epsilon_1}}{\bar{\epsilon}'_1}\biggr)\,
\biggl\{
\langle\phi|4\,\pi\,\delta^3_\lambda(r)\,
\frac{1}{(E-H)'}\,
4\,\pi\,\delta^3_\lambda(r)|\phi\rangle+
\frac{1}{2}\,\langle\phi|\nabla^2\,4\,\pi\,\delta^3_\lambda(r)|\phi\rangle
\biggr\}\,.
\end{equation}
Both terms in above braces have already been calculated in the context
of positronium energy levels in \cite{krp3}
\begin{eqnarray}
\langle\phi|4\,\pi\,\delta^3_\lambda(r)\,
\frac{1}{(E-H)'}\,
4\,\pi\,\delta^3_\lambda(r)|\phi\rangle &=& 
-\frac{8}{n^3}\,\biggl[\frac{\lambda}{2}+2\,\ln\frac{\lambda}{3}
+8\,\ln\frac{3}{4}- \frac{3}{2}+\frac{2}{n} \nonumber \\ &&
+2\bigl(\ln(n)-\Psi(n)-C)\biggr]\,, \label{26} \\
\langle\phi|\nabla^2\,4\,\pi\,\delta^3_\lambda(r)|\phi\rangle &=&
-\frac{8}{n^3}\,\biggl[-\frac{1}{n^2}+\lambda-4+6\,
\ln\frac{3}{4}\biggr]\,, \label{27}
\end{eqnarray}
with $n=1$ in our case.
$B_H$ is the difference between $B$ and $B_L$. In this difference only
large electron momenta contribute, therefore it could be obtained
in the scattering amplitude approximation, in the same way as $P_2$
and $P_3$ in a simple example in the previous section.
The result is
\begin{eqnarray}
B_H = {\cal E}\,\frac{4}{9} &\biggl [&
8 + 5\,{\pi }^2 - \ln \biggl(\frac{{\bar{\epsilon}_1}}{{\bar{\epsilon}'_1}}\biggr) + 
   2\,\lambda\,\ln \biggl(\frac{{\bar{\epsilon}_1}}{{\bar{\epsilon}'_1}}\biggr) - 
   50\,\ln (2)\,\ln \biggl(\frac{{\bar{\epsilon}_1}}{{\bar{\epsilon}'_1}}\biggr) + 
   18\,\ln (3)\,\ln \biggl(\frac{{\bar{\epsilon}_1}}{{\bar{\epsilon}'_1}}\biggr) 
   \nonumber \\ && + 
   {\ln \biggl(\frac{{\bar{\epsilon}'_1}}{{\bar{\epsilon}_2}}\biggr)}^2 + 
   4\,\ln \biggl(\frac{{\bar{\epsilon}_1}}{{\bar{\epsilon}'_1}}\biggr)\,
    \ln \biggl(\frac{\lambda}{{\sqrt{{\bar{\epsilon}_2}}}}\biggr) \biggr]\,.
\end{eqnarray}
The complete $B$ term is
\begin{eqnarray}
B = {\cal E}\,\frac{4}{9} & [&
 8 + 5\,{\pi }^2 + 3\,\ln ({\bar{\epsilon}_1}) - 
   6\,\ln (2)\,\ln ({\bar{\epsilon}_1}) - 
   2\,\ln ({\bar{\epsilon}_1})\,\ln ({\bar{\epsilon}_2}) + 
   {\ln ({\bar{\epsilon}_2})}^2 - 3\,\ln ({\bar{\epsilon}'_1}) 
   \nonumber \\ &&+ 
   6\,\ln (2)\,\ln ({\bar{\epsilon}'_1}) + {\ln ({\bar{\epsilon}'_1})}^2]\,.
\end{eqnarray} 
We can now go back to Eq. (\ref{24}) for the second derivative of ${\cal E}_L$
which is a sum of $A$ and $B$
\begin{eqnarray}
\epsilon_2\,\frac{\partial {\cal E}_L}{\partial \epsilon_2} &=&
{\cal E}\,\frac{4}{9}\, [
8 + \frac{3\,N}{2} + 5\,{\pi }^2 + 3\,\ln ({\bar{\epsilon}_1}) - 
   6\,\ln (2)\,\ln ({\bar{\epsilon}_1}) - 
   2\,\ln ({\bar{\epsilon}_1})\,\ln ({\bar{\epsilon}_2}) + {\ln ({\bar{\epsilon}_2})}^2]\,.
\end{eqnarray}
The expression for ${\cal E}_L$ which matches both derivatives is:
\begin{eqnarray}
{\cal E}_L(\bar{\epsilon}_1,\bar{\epsilon}_2) = {\cal E}
& \biggl[ &  \frac{2\,N\,\ln ({\bar{\epsilon}_1})}{3} + 
   \frac{32\,\ln ({\bar{\epsilon}_2})}{9} + 
   \frac{2\,N\,\ln ({\bar{\epsilon}_2})}{3} + 
   \frac{20\,{\pi }^2\,\ln ({\bar{\epsilon}_2})}{9} + 
   \frac{4\,\ln ({\bar{\epsilon}_1})\,\ln ({\bar{\epsilon}_2})}{3} 
   \nonumber \\ && - 
   \frac{8\,\ln (2)\,\ln ({\bar{\epsilon}_1})\,\ln ({\bar{\epsilon}_2})}{3} - 
   \frac{4\,\ln ({\bar{\epsilon}_1})\,{\ln ({\bar{\epsilon}_2})}^2}{9} + 
   \frac{4\,{\ln ({\bar{\epsilon}_2})}^3}{27}\biggr]\,.
\end{eqnarray}
The constant term (no logs) is not included here.
$E_L$ as shown in Fig. \ref{fig1} is integrated over the region
which is a combination of three rectangles:
\begin{equation}
E_L = {\cal E}_L\biggl(\frac{\epsilon'_1}{\alpha^2},\frac{\epsilon_2}{\alpha^2}\biggr)+
      {\cal E}_L\biggl(\frac{\epsilon'_2}{\alpha^2},\frac{\epsilon_1}{\alpha^2}\biggr)-
      {\cal E}_L\biggl(\frac{\epsilon_1}{\alpha^2},\frac{\epsilon_2}{\alpha^2}\biggr)\,. 
      \label{32}
\end{equation}
  
\section{Contribution $E_M$}
In the one-loop case, contribution to energy, coming from
photon energies $k^0>\epsilon$ is
\begin{eqnarray}
\delta E &=& \langle\phi|V|\phi\rangle\,, \\
V(\epsilon) &=&\alpha^2\,\delta^3(r)\biggl[
\frac{10}{9}-\frac{4}{3}\,\ln(2\,\epsilon)\biggr]\,. \label{34}
\end{eqnarray}
$E_M$ is a $V$ correction to the Bethe log:
\begin{equation}
E_M = \frac{2\,\alpha}{3\,\pi}\,\delta_{V(\epsilon_1)}\,
\int_0^{\epsilon_2} d\omega\,\omega\,
\langle\phi|p^i\,\frac{1}{E-(H+\omega)}\,p^i|\phi\rangle\,.
\end{equation}
It has the same form as Eq. (\ref{22}), so after symmetrization
$\epsilon_1\leftrightarrow\epsilon_2$ it is:
\begin{equation}
E_M = \frac{{\cal E}}{2}\,
\biggl(\frac{10}{9}-\frac{4}{3}\,\ln(2\,\epsilon'_1) \biggr)\,
\biggl[
-\frac{2}{3}\,\ln^2\frac{\epsilon_2}{\alpha^2}+
2\,(1-2\,\ln 2)\,\ln\frac{\epsilon_2}{\alpha^2}+N\biggr]+
(\epsilon_1\leftrightarrow\epsilon_2)\,. \label{36}
\end{equation}

\section{Contribution $E_F$}
$E_F$ is the two-loop contribution with regularized Coulomb interaction
and with both photon energies limited from below by $\epsilon$.
It is a sum of three terms 
\begin{equation}
E_F = E_F^1+E_F^2+E_F^3\,, \label{37}
\end{equation}
defined and calculated as follows.
$E_F^1$ is a second order correction coming from $V(\epsilon_1)$
and $V(\epsilon_2)$ with $V$ defined in (\ref{34}), here additionally with
$\lambda$-regularization
\begin{equation}
E_F^1 = \langle \phi| V(\epsilon_1)\,\frac{1}{(E-H)'}\,V(\epsilon_2)
        |\phi\rangle\,.
\end{equation}
The corresponding matrix element is given in Eq. (\ref{26}), so $E_F^1$ becomes
\begin{equation}
E_F^1 = \frac{{\cal E}}{16}\,
\biggl(\frac{10}{9}-\frac{4}{3}\,\ln(2\,\epsilon_1)\biggr)\,
\biggl(\frac{10}{9}-\frac{4}{3}\,\ln(2\,\epsilon_2)\biggr)\, 
\biggl(-4\,\lambda-16\,\ln\frac{\lambda}{3}\,-4-64\,\ln\frac{3}{4}\biggr)\,. \label{39}
\end{equation}
One needs only $\ln\lambda$ term, since others do not give $\ln\alpha$.
$E_F^2$ is the contribution from electron formfactors
$F'_1$ and $F_2$ at $q^2=0$ on relativistic (Dirac) wave function.
We know it from the one-loop case that for vacuum-polarization
 $A_{61}= A_{40}/2$.
The same holds for two-loop contribution, thus we have
\begin{equation}
E_F^2 = {\cal E}\, \ln\alpha^{-2}\, \frac{B_{40}}{2}\,.
\end{equation}
Diagrams with closed fermion loop are
automatically included
in the above formula. Other contributions coming from these diagrams
are calculated in Section VII.

$E_F^3$ is the contribution from $F''_1$ and $F'_2$ calculated 
with nonrelativistic wave functions. It leads to the matrix element
$\langle\phi|\nabla^2\,\delta^3(r)_\lambda|\phi\rangle$ which does
not lead to $\ln\lambda$. Hence, it does not contribute to $\ln\alpha$. 

\section{Contribution $E_H$}
$E_H$ is the contribution obtained from the two--loop three--photon exchange
forward scattering amplitude. It requires subtractions of terms,
contributing to Lamb shift at lower orders.
After subtractions it is finite and depends on 
$\epsilon_1,\epsilon_2$ and $\Lambda = \lambda\,\alpha$.
When combined with $E_L$ and  $E_F$, the dependence on 
$\epsilon_1,\epsilon_2$ and $\Lambda$ should cancel out.
Having this in mind, the $\ln\alpha$ contribution could be obtained
by the replacement $\lambda\rightarrow 1/\alpha$ in $E_F^1$ in Eq. (\ref{39}).
However, the constant term $B_{60}$ requires complete calculation of $E_H$,
which we think is the most difficult of the contributions.  

\section{Diagrams with closed fermion loop}
There is a small logarithmic contribution coming from diagrams
with a closed fermion loop. They are partially included in $E_F^2$.
Two other contributions $E_{VP}^1, E_{VP}^2$ are the following. 
The second order correction coming from the one-loop vacuum polarization is
\begin{equation}
E_{VP}^1 = {\cal E}\,
\biggl(-\frac{4}{15}\biggr)^2\,
\langle\phi|\delta^3_\lambda(r)\,\frac{1}{(E-H)}\,
\delta^3_\lambda(r)|\phi\rangle
\rightarrow {\cal E}\,
  \biggl(\frac{4}{15}\biggr)^2\,\ln\,\alpha\,. \label{41}
\end{equation}
The second contribution $E_{VP}^2$ is electron self--energy in the 
Coulomb potential including vacuum polarization correction.
It is calculated in the similar way, as previous corrections.
One splits it into three parts
\begin{equation}
E^2_{VP} = C_L+C_M+C_H\,.
\end{equation}
$C_L$ is a v.p. correction $ V=-(4/15) \,\delta^3(r)$ to the Bethe log:
\begin{eqnarray}
C_L &=& \frac{2\,\alpha}{3\,\pi}\,\delta_V\,\int_0^\epsilon\,d\omega\,\omega\,
\langle\phi|p^i\,\frac{1}{E-(H+\omega)}\,p^i|\phi\rangle\\
&=& {\cal E}\,\biggl(-\frac{4}{15}\biggr)\,
\biggl(-\frac{2}{3}\,\ln^2\frac{\epsilon}{\alpha^2}+
2\,(1-2\,\ln 2)\,\ln\frac{\epsilon}{\alpha^2}+N\biggr)\,.
\end{eqnarray} 
$C_M$ is a second order correction coming from self--energy and v.p.
\begin{eqnarray}
C_M &=& 2\,\biggl(\frac{\alpha}{\pi}\biggr)^2\,
\biggl(\frac{10}{9}-\frac{4}{3}\,\ln 2\,\epsilon\biggr)\,
\biggl(-\frac{4}{15}\biggr)\,
\langle\phi|\delta^3_\lambda(r)\,\frac{1}{(E-H)}\,
\delta^3_\lambda(r)|\phi\rangle \\
 &\rightarrow& 2\,{\cal E}\,
\biggl(\frac{10}{9}-\frac{4}{3}\,\ln 2\,\epsilon\biggr)\,
\biggl(-\frac{4}{15}\biggr)\,\ln\alpha\,.
\end{eqnarray}
$C_H$ is given by the scattering amplitude. Since we
calculate only the logarithmic part, instead of calculating $B_H$
we replaced $\ln\lambda$ by $-\ln\alpha$ in the equation above.
The logarithmic part of electron self--energy in the 
Coulomb potential including vacuum polarization correction is
\begin{equation}
E_{VP}^2 = {\cal E}\,\frac{4}{15}\,\biggl[
\frac{2}{3}\,\ln^2\alpha^{-2}+4\biggl(\frac{2}{9}+\ln2\biggr)\,
\ln\alpha^{-2}\biggr]\,. \label{47}
\end{equation}
This completes the treatment of two-loop logarithmic correction

\section{Summary} 
The sum of all logarithmic terms in Eqs. 
(\ref{32},\ref{36},\ref{37},\ref{41},\ref{47}) is
\begin{eqnarray}
B_{63} &=&-\frac{8}{27} \hspace{20mm}= -0.296296\,,\\
B_{62} &=& \frac{104}{135} - \frac{16\,\ln 2}{9}\hspace{3.5mm}=-0.461891\,, \\
B_{61} &=& \frac{39751}{10800} + \frac{4\,N}{3} + 
\frac{55\,\pi^2}{27} - \frac{616\,\ln 2}{135} + 
\frac{3\,\pi^2\,\ln 2}{4} + \frac{40\,\ln^2 2}{9} - 
\frac{9\,\zeta(3)}{8} \\&&\hspace{30mm}= 50.309654\,.\nonumber
\end{eqnarray}
First of all the result for $B_{61}$ is surprisingly large, and reverses
the sign of the overall logarithmic contribution. $B_{63}$ agrees with
the result obtained first in \cite{sk1}.
However, as it was pointed out by Yerokhin \cite{yer}, the loop-by-loop diagram
is the source of additional terms, which were not accounted for
in the calculation in \cite{sk1}. 
An additional result of this work is the state dependence of $B$
coefficients
which is obtained from $n$-dependence of matrix elements in 
Eqs. (\ref{20},\ref{26},\ref{27})
\begin{eqnarray}
B_{62}(n)&=& B_{62}+\frac{16}{9}\,
\biggl(\frac{3}{4} + \frac{1}{4\,n^2} - \frac{1}{n} - 
\ln(n)+ \Psi(n)+C\biggr)\,,\\
B_{61}(n)&=& B_{61}+\frac{4}{3}\,\bigl(N(n)-N\bigr) + 
\biggl(\frac{304}{135} - \frac{32}{9}\ln(2)\biggr)\,
\biggl(\frac{3}{4} + \frac{1}{4\,n^2} - \frac{1}{n} - 
\ln(n)+ \Psi(n)+C\biggr)\,.
\end{eqnarray}
$n$-dependence of $B_{62}$  agrees with the former result in \cite{sk2}
(apart from the misprint in the overall sign there). 
$B_{61}$ depends on $N$-coefficient, the Dirac delta correction to Bethe logs, 
which has not been calculated yet for other states than 1S, therefore
its complete state dependence is unknown. However, one may expect 
to a good approximation $N$ is independent of $n$, as it is for Bethe logs. 

Because of the large value of $B_{61}$ theoretical predictions
for hydrogen Lamb shift are going to be changed.
The total logarithmic contribution is 16.9 kHz for the 1S state, 
compared to the previous one, based only on $B_{63}$ -28.4 kHz.
Theoretical predictions for Lamb shift in hydrogen
with  proton radius $r_p = 0.862(12)$ fm from \cite{rp}, 
using recent updates: 
analytical calculations of the three-loop contribution by Melnikov and 
Ritbergen in \cite{mel} and direct numerical calculation
of one-loop self-energy by Jentschura {\em et al.} in \cite{yen}  are
(see details in the appendix)
\begin{eqnarray}
E_L(1S)_{\rm th} &=& 8\,172\,816(10)(32)\, {\rm kHz}, \\
E_L(2S-2P_{1/2})_{\rm th} &=&1\,057\,842(1)(4)\, {\rm kHz}, 
\end{eqnarray}
where we assumed for $B_{60} = 0 \pm 100$, which gives the first uncertainty.
For $P$-states we neglect $B$-terms completely.
The second uncertainty comes from the proton charge radius.
Since it dominates the theoretical error,
we emphasize the importance of the muonic-hydrogen measurement,
from which $r_p$ could be precisely obtained.  
Current theoretical predictions agrees well with the most precise
experimental values:
\begin{eqnarray}
E_L(1S)_{\rm exp} &=& 8\,172\,837(22)\,{\rm kHz} \cite{gar,paris}, \\
E_L(2S-2P_{1/2})_{\rm exp} &=&1\,057\,845(9)\, {\rm kHz} \cite{lund}, \\
E_L(2S-2P_{1/2})_{\rm exp} &=&1\,057\,842(12)\, {\rm kHz} \cite{hag}. \\
\end{eqnarray}
Due to large uncertainty and ambiguities with the proton charge radius,
one may regard the Lamb measurement as a determination of $r_p$.
In this way, from 1S Lamb shift, one obtains:
\begin{equation}
r_p = 0.869(12)\,{\rm fm}\,.
\end{equation} 
Logarithmic two-loop corrections significantly alter theoretical predictions
for the Lamb shift in the single ionized helium as well. 
The current theoretical value is
\begin{equation}
E_L(2S-2P_{1/2})_{\rm th} = 14\,041.57(8)\,{\rm MHz}\,.
\end{equation}
It does not agree with both: the experimental value from \cite{drake1} and 
the recent update in \cite{drake2} respectively:
\begin{eqnarray}
E_L(2S-2P_{1/2})_{\rm exp} &=& 14\,042.52(16)\,{\rm MHz}\,, \\
E_L(2S-2P_{1/2})_{\rm exp} &=& 14\,041.13(17)\,{\rm MHz}\,.
\end{eqnarray}
One may wonder about $B_{60}$ and 
further higher order terms, keeping in mind the large value of $B_{61}$.
There are two possible and complementary undergoing projects: direct calculation
of this term or numerical calculation of complete two-loop diagrams
with Dirac-Coulomb propagators. While the second would be the best way,
the numerical accuracy might be limited at small $Z$, such as $Z=1$.
In the direct calculation of $B_{60}$ one has to consider
three points: two-loop Bethe logs with $\epsilon$ cut-offs, 
two-loop scattering amplitude with the photon 
mass $\mu$, and the transition terms between 
$\epsilon$ and $\mu$. This project seems to be achievable
using the methods developed for $B_{50}$, positronium decay rate
and the one applied here.

\section*{Acknowledgments}
I gratefully acknowledge interesting discussions and helpful comments 
from Jonathan Sapirstein. I wish to thank M. Eides for inspiration.
This work was supported by Polish Comittee
for Scientific Research under Contract No. 2P03B 057 18.

\appendix

\section{Formulas for calculations of Lamb shift}
\noindent In the calculation of hydrogen and helium Lamb shift
we use the following physical constants:
\begin{eqnarray}
R               &=& 10973731.568516(84)\,{\rm m}^{-1}    \,, \nonumber \\
c               &=& 299792458\,{\rm m\,s}^-1              \,,\nonumber\\
\alpha^{-1}     &=& 137.03599958(50)       \,,\nonumber\\
\frac{m_p}{m_e} &=& 1836.1526675(39)       \,,\nonumber\\   
\frac{m_\alpha}{m_e}  &=& 7294.299508(16)   \,,\nonumber\\      
r_p                   &=& 0.862(12)\,{\rm fm} \,,\nonumber\\   
r_\alpha              &=& 1.673(1)\,{\rm fm}\,.
\end{eqnarray}
In  general, Lamb shift in light hydrogen like systems is a sum of nonrecoil, 
recoil and the proton structure contributions.
In the nonrecoil limit, known terms are:
\begin{eqnarray}
E_L &=& m\,\frac{\alpha\,(Z\,\alpha)^4}{\pi\,n^3}\left(\frac{\mu}{m}\right)^3
\bigl\{
A_{40}+A_{41}\,L+(Z\,\alpha)\,A_{50}+(Z\,\alpha)^2\,\bigl[
A_{62}\,L^2+A_{61}\,L+A_{60}(Z\,\alpha)\bigr]\nonumber \\ &&
+\frac{\alpha}{\pi}\,\bigl[
B_{40}+(Z\,\alpha)\,B_{50}+(Z\,\alpha)^2\,
\bigl(B_{63}\,L^3+B_{62}\,L^2+B_{61}\,L+B_{60}(Z\,\alpha)\bigr)\bigr]
\nonumber \\ &&
+\left(\frac{\alpha}{\pi}\right)^2\,C_{40} \bigr\}\,, \label{A2}
\end{eqnarray}
where $\mu$ is a reduced mass, $m=m_e$, and 
$L=\ln\bigl[m/(\mu\,(Z\,\alpha)^2)\bigr]$.
Most of these coefficients could be find in any review, such as 
\cite{sap1} or \cite{eides2}. The recent result is the  direct numerical calculations
of one-loop self-energy, which gives for hydrogen  $(Z=1)$
\begin{eqnarray}
A_{60}(1S,\alpha) &=& -30.29024 +\biggl[-0.6187 +
                      \biggl(\frac{19}{45}-\frac{\pi^2}{27}\biggr)\biggr]\,, \nonumber \\
A_{60}(2S,\alpha) &=& -31.18515 +\biggl[-0.8089 +
                      \biggl(\frac{19}{45}-\frac{\pi^2}{27}\biggr)\biggr]\,, \nonumber \\
A_{60}(2P_{1/2},\alpha) &=& -0.9735-0.0640\,, 
\end{eqnarray}
and for He$^+$ $(Z=2)$
\begin{eqnarray}
A_{60}(2S,2\,\alpha) &=& -30.64466+\biggl[-0.7961+
                       \biggl(\frac{19}{45}-\frac{\pi^2}{27}\biggr)\biggr]\,,\nonumber \\
A_{60}(2P_{1/2},2\,\alpha) &=& -0.94940-0.0638\,,  
\end{eqnarray}
where the second term is the vacuum polarization \cite{mohr}. Another  recent result is 
the analytical calculation of three-loop contribution in \cite{mel}. Together with 
the previously known vacuum polarization and anomalous magnetic moment it amounts to
\begin{equation}
C_{40} = 0.417508\,.
\end{equation}
In this work we calculate all logarithmic two-loop corrections for S-states.
However, for P-state only $B_{62}$ is known. For this reason in 
the theoretical predictions for hydrogen and helium we totally neglect
higher order two loop corrections for $P$ states, but included $B_{40}$ only.
We neglect also dependence of $N$ in Eq. (\ref{20}) on principal quantum number $n$,
since $N$ has not yet been calculated for $n \neq 1$. 
Recoil corrections, not included in Eq. (\ref{A2}) sum to
\begin{eqnarray}
\delta E &=& \frac{\mu^3}{m\,M}\,\frac{(Z\,\alpha)^5}{\pi\,n^3}\,\biggl\{
\frac{1}{3}\,\delta_{l0}\,\ln(Z\,\alpha)^{-2}-\frac{8}{3}\,\ln k_0(n,l) 
+\frac{14}{3}\,\delta_{l0}\,\biggr[\ln\biggl(\frac{2}{n}\biggr)+
\Psi(n)+C+\frac{1}{2\,n}+1\biggr]
 \nonumber \\ && -\frac{1}{9}\,\delta_{l0}
-\frac{2}{M^2-m^2}\,\delta_{l0}\,\biggl[M^2\,\ln\biggl(\frac{m}{\mu}\biggr)-
m^2\,\ln\biggl(\frac{M}{\mu}\biggr)\biggr]
-\frac{7}{3}\,\frac{1-\delta_{l0}}{l\,(l+1)\,(2\,l+1)}
\biggr\} \nonumber \\&&
-\alpha\,\frac{(Z\,\alpha)^5}{n^3}\,\frac{m^2}{M}\,\delta_{l0}\,[1.364\,49(2)]
+\frac{(Z\,\alpha)^6}{n^3}\,\frac{m^2}{M}\,D_{60}\,,
\end{eqnarray}
where
\begin{eqnarray}
D_{60}(nS_{1/2}) &=& 4\,\ln(2)-\frac{7}{2}\,,\nonumber \\
D_{60}(l\geq 1) &=& \biggl[ 3-\frac{l\,(l+1)}{n^2}\biggr]\,
\frac{2}{(4\,l^2-1)(2\,l+3)}\,.
\end{eqnarray}
The finite charge distribution of the nucleus and its self-energy give corrections:
\begin{eqnarray}
\delta E &=& \frac{2}{3\,n^3}\,(Z\,\alpha)^4\,\mu^3\,r^2\,\delta_{l0}+
\frac{4}{3\,\pi\,n^3}\,\frac{\mu^3}{M^2}\,(Z^2\,\alpha)\,(Z\,\alpha)^4\,
\left[\ln\biggl(\frac{M}{\mu\,(Z\,\alpha)^2}\biggr)\,\delta_{l0}-\ln k_0(n,l)\right]\,.
\end{eqnarray}
In the theoretical predictions, presented in this paper we have neglected higher order
proton structure corrections and higher order recoil corrections, which at present
are negligible.

\end{document}